\documentclass[11pt, oneside]{article}   	%
\usepackage{geometry}                		%
\usepackage{graphicx}				%
\usepackage{amssymb}
\usepackage{physics}
\usepackage[colorlinks=false, urlcolor=blue, linkcolor=red]{hyperref}
\usepackage{cite}
\usepackage[symbol]{footmisc}

\title{\bf Reply to ``Comment on work of Yang and Nevels `Direct, analytic solution for the electromagnetic vector potential 
in any gauge' (arXiv:2507.02104)'' by Onoochin (arXiv:2507.08042)}
\author
{Kuo-Ho Yang \\
Department of Engineering and Physics, St.~Ambrose University, Davenport, IA 52803 \\
E-mail: yangkuoho@sau.edu\vspace{2mm}\\
Robert D. Nevels\\
Department of Electrical and Computer Engineering, Texas A\ \&\ M University,\\
College Station, TX 77843
}
\date{}
\begin{document}
\maketitle

\begin{abstract}
Because of a difference between his and our results for the Coulomb-gauge vector potential, Onoochin (arXiv:2507.08042) suspected that we might have used some (so far unidentified) mathematically illegal operations in our paper (arXiv:2507.02104).  In this note, we present four methods to prove the validity of our results in dispute.  Onoochin's method for calculating his Coulomb-gauge potential is among our four methods used in the proofs.
\end{abstract} \vspace{7mm}

In a recent note\cite{Onoochin-2025}, Onoochin calculated the Coulomb-gauge vector potential for a point-charge moving at a constant speed on a straight line and discovered that his Coulomb-gauge vector potential appeared to differ from our Coulomb-gauge vector potential in Ref.~\cite{Yang-Nevels-2025}.  Based on this discrepancy, he suspected that we might have used some (so far unidentified) ``{\em mathematically illegal operations}'' to derive results in Ref.~\cite{Yang-Nevels-2025}.  In this note, we present four different methods to prove the validity of our results for an arbitrary time-dependent charge-current distribution.  And, we do all our proofs analytically.  Onoochin's method for calculating his Coulomb-gauge vector potential is among our four methods presented here.

We consider localized charge and current densities, $\rho ({\bf r}, t)$ and ${\bf J} ({\bf r}, t)$, which are turned on at $t_0$.  The electric field ${\bf E}$, the magnetic field ${\bf B}$, and Maxwell's equations for potentials ${\bf A}$ and $\Phi$ are (in Gaussian units):
\begin{equation}
{\bf E}({\bf r}, t) = - \grad \Phi({\bf r}, t) - {1 \over c} {\partial {\bf A}({\bf r}, t) \over \partial t},
\qquad
{\bf B}({\bf r}, t) = \grad \times {\bf A}({\bf r}, t),
\label{eq-A1}
\end{equation}
\begin{equation}
\grad^2 \Phi ({\bf r}, t) + {1 \over c} {\partial \over \partial t} [ \grad \cdot {\bf A}({\bf r}, t) ] 
= - 4 \pi \rho ({\bf r}, t),
\label{eq-A2}
\end{equation}
\begin{equation}
\left( \grad^2 - {1 \over c^2} {\partial^2 \over \partial t^2}  \right) {\bf A}({\bf r}, t)  
= - { 4 \pi \over c } {\bf J}({\bf r}, t) 
  + \grad \left( \grad \cdot {\bf A}({\bf r}, t) + {1 \over c} {\partial \Phi ({\bf r}, t) \over \partial t} \right).
\label{eq-A3}
\end{equation}

The velocity gauge with parameter $v$ has the gauge condition\cite{Yang-Nevels-2025}:
\begin{equation}
 \grad \cdot {\bf A}^{(v)} + {c \over v^2} {\partial \Phi^{(v)} \over \partial t} = 0.
 \label{eq-A4}
\end{equation}
We choose the velocity gauge in this note because the questions raised in Ref.~\cite{Onoochin-2025} pertain to this gauge.
Thus, eqs.~(\ref{eq-A2})-(\ref{eq-A3}) become:
\begin{equation}
\left( \grad^2 - {1 \over v^2} {\partial^2 \over \partial t^2} \right) \Phi^{(v)} = - 4\pi \rho,
\label{eq-A5}
\end{equation}
\begin{equation}
\left( \grad^2 - {1 \over c^2} {\partial^2 \over \partial t^2} \right) {\bf A}^{(v)}
= - {4 \pi \over c} {\bf J} + c \left( - {1 \over v^2} + {1 \over c^2} \right) \left( {\partial \over \partial t} \grad \Phi^{(v)} \right).
\label{eq-A6}
\end{equation}
To solve eq.~(\ref{eq-A6}), we write ${\bf A}^{(v)} = {\bf A}^{(v)}_1 + {\bf A}^{(v)}_2$ where ${\bf A}^{(v)}_1$ and ${\bf A}^{(v)}_2$ are solutions of their respective equations:
\begin{equation}
\left( \grad^2 - {1 \over c^2} {\partial^2 \over \partial t^2} \right) {\bf A}^{(v)}_1
= - {4 \pi \over c} {\bf J},
\label{eq-A7}
\end{equation}
\begin{equation}
\left( \grad^2 - {1 \over c^2} {\partial^2 \over \partial t^2} \right) {\bf A}^{(v)}_{2}
= c \left( - {1 \over v^2} + {1 \over c^2} \right) \left( {\partial \over \partial t} \grad \Phi^{(v)} \right).
\label{eq-A8}
\end{equation}
The solutions for ${\bf A}^{(v)}_1$ and ${\bf A}^{(v)}_2$ were given in Ref.~\cite{Yang-Nevels-2025}, eqs.~(6), (8), and (15):
\begin{equation}
{\bf A}^{(v)}_1({\bf r}, t) = {\bf A}_c ({\bf r}, t)= {1 \over c} \int {{\bf J}({\bf r}', t - R/c) \over R} d^3r',
\qquad \qquad
R = |{\bf r} - {\bf r}'|,
\label{eq-A9}
\end{equation}
\begin{equation}
{\bf A}^{(v)}_2 ({\bf r}, t) = c \grad \int \left( \Phi_c ({\bf r}, t) - \Phi^{(v)} ({\bf r}, t) \right) dt,
\label{eq-A10}
\end{equation}
\begin{equation}
\Phi^{(v)} ({\bf r}, t) = \int {\rho({\bf r}', t - R/v) \over R} d^3r'
= \int G({\bf r}, t |v| {\bf r}', t') \rho({\bf r}', t') d^3r'dt',
\label{eq-A11}
\end{equation}
\begin{equation}
G({\bf r}, t |v| {\bf r}', t') = {\delta(t - R/v - t') \over R}, 
\qquad \qquad
R = |{\bf r} - {\bf r}'|,
\label{eq-A12}
\end{equation}
\begin{equation}
\left( \grad^2 - {1 \over v^2} {\partial^2 \over \partial t^2} \right) G({\bf r}, t |v| {\bf r}', t') 
= - 4 \pi \delta({\bf r} - {\bf r}') \delta(t - t'),
\ 
\label{eq-A13}
\end{equation}
and $\Phi_c = \left[ \Phi^{(v)} \right]_{v = c}$ is the $c$-retarded scalar potential from the charge density $\rho$ (by substituting $c$ for $v$ in eq.~(\ref{eq-A11})).
In the following, we present four different methods to prove that the vector potential ${\bf A}^{(v)}_2$ in eq.~(\ref{eq-A10}) is a mathematically correct solution to eq.~(\ref{eq-A8}).

\begin{flushleft}
{\bf Method 1}:
\end{flushleft}

Here, we begin our proofs by applying $(\grad^2 - c^{-2} \partial^2/\partial t^2)$ from the left to eq.~(\ref{eq-A10}) and then use eq.~(\ref{eq-A5}) to get:
\begin{eqnarray}
 \left( \grad^2  - {1 \over c^2} {\partial^2 \over \partial t^2} \right){\bf A}^{(v)}_2 
 = c \grad \int \left[  \left( \grad^2 \Phi_c - {1 \over c^2} {\partial^2 \over \partial t^2}\Phi_c \right) 
   - \left( \grad^2  \Phi^{(v)} - {1 \over c^2} {\partial^2 \over \partial t^2}\Phi^{(v)} \right) \right] dt
\nonumber
\\
= c \grad \int \left[ \left( -4\pi \rho \right)
   -  \left( - 4\pi \rho +  {1 \over v^2} {\partial^2 \over \partial t^2}  \Phi^{(v)}  - {1 \over c^2} {\partial^2 \over \partial t^2}\Phi^{(v)}  \right) \right]dt
\qquad \qquad \qquad \qquad \qquad 
\nonumber
\\
= c \grad \int \left( - {1 \over v^2} + {1 \over c^2} \right) {\partial^2 \over \partial t^2}\Phi^{(v)} dt
= c \left(- {1 \over v^2} + {1 \over c^2}  \right)  \left({\partial \over \partial t}  \grad \Phi^{(v)} \right).
\qquad \qquad \qquad \quad \quad \,
\label{eq-A14}
\end{eqnarray}
Thus, we have proved eq.~(\ref{eq-A10}) by the method of simple substitution.  This simple test should always be done first before any conclusions are reached.

\begin{flushleft}
{\bf Method 2}:
\end{flushleft}

We now solve ${\bf A}^{(v)}_2$ directly from eq.~(\ref{eq-A8}) by first differentiating both sides of the equation by time $t$ to get
\begin{eqnarray}
\left( \grad^2 - {1 \over c^2} {\partial^2 \over \partial t^2} \right) {\partial {\bf A}^{(v)}_{2} \over \partial t}
= \left( - {1 \over v^2} {\partial^2 \over \partial t^2} + {1 \over c^2} {\partial^2 \over \partial t^2} \right) \left( c \grad \Phi^{(v)} \right)
\nonumber
\\
=\left[ \left(\grad^2 - {1 \over v^2} {\partial^2 \over \partial t^2} \right) - \left( \grad^2 
  - {1 \over c^2} {\partial^2 \over \partial t^2} \right) \right] \left( c \grad \Phi^{(v)} \right)
\qquad \qquad 
\nonumber
\\
= - 4 \pi c \grad \rho
- \left( \grad^2 - {1 \over c^2} {\partial^2 \over \partial t^2} \right) \left( c \grad \Phi^{(v)} \right).
\qquad \qquad \qquad \quad \ \ \,
\label{eq-A15}
\end{eqnarray}
If we move the last term to the LHS of the equation, then
\begin{equation}
\left( \grad^2 - {1 \over c^2} {\partial^2 \over \partial t^2} \right)
\left( {\partial {\bf A}^{(v)}_{2} \over \partial t} + c\grad \Phi^{(v)} \right)
= - 4 \pi c \grad \rho.
\label{eq-A16}
\end{equation}
The equation can be solved by the $c$-retarded Green's function:
\begin{equation}
{\partial {\bf A}^{(v)}_{2}({\bf r}, t) \over \partial t} + c\grad \Phi^{(v)}({\bf r}, t)
= c\grad \int {\rho({\bf r}', t - R/c) \over R} d^3r' = c\grad \Phi_c({\bf r}, t),
\label{eq-A17}
\end{equation}
where $R = |{\bf r} - {\bf r}'|$. The above result obviously leads to the solution of ${\bf A}^{(v)}_2$ in eq.~(\ref{eq-A10}).
\begin{flushleft}
{\bf Method 3}:
\end{flushleft}

In our third method we obtain ${\bf A}^{(v)}_2$ by integrating eq.~(\ref{eq-A8}) and using eqs.~(\ref{eq-A11})-(\ref{eq-A13}):
\begin{eqnarray}
{\bf A}^{(v)}_2 ({\bf r}, t) = {c \over 4\pi} \left( {1 \over v^2} - {1 \over c^2} \right)  \grad {{\partial \over \partial t}}
\int G({\bf r}, t |c| {\bf r}'', t'') \Phi^{(v)}({\bf r}'', t'') d^3r''dt''
\qquad \ 
\nonumber
\\
= {c \over 4\pi} \left( {1 \over v^2} - {1 \over c^2} \right)  \grad {{\partial \over \partial t}}
\int G({\bf r}, t |c| {\bf r}'', t'') G({\bf r}'', t'' |v| {\bf r}', t' ) \rho({\bf r}', t') d^3r''dt'' d^3r'dt' 
\nonumber
\\
= {c \over 4\pi} \left( {1 \over v^2} - {1 \over c^2} \right)  \grad {{\partial \over \partial t}}
\int G({\bf r}, t |c|v| {\bf r}', t') \rho({\bf r}', t') d^3r'dt',
\qquad \qquad \qquad \qquad \quad \ 
\label{eq-A18}
\end{eqnarray}
where (see eqs.~(3.33) and (3.34) of Ref.~\cite{Yang-2005})
\begin{equation}
G({\bf r}, t |c|v| {\bf r}', t') = \int G({\bf r}, t |c| {\bf r}'', t'') G({\bf r}'', t'' |v| {\bf r}', t') d^3r''dt'',
\label{eq-A19}
\end{equation}
\begin{equation}
\left( \grad^2 - {1 \over v^2} {\partial^2 \over \partial t^2} \right) \left( \grad^2 - {1 \over c^2} {\partial^2 \over \partial t^2} \right)
G({\bf r}, t |c|v| {\bf r}', t') = (- 4\pi)^2 \delta({\bf r} - {\bf r}') \delta(t - t').
\label{eq-A20}
\end{equation}
The above result in eq.~(\ref{eq-A18}) certainly does {\em not} look like ${\bf A}^{(v)}_2$ in eq.~(\ref{eq-A10}).  

Thus, we need to simplify eq.~(\ref{eq-A18}) into a form that can be physically interpreted as a wave propagating directly from a source, such as ${\bf A}_c$ in eq.~(\ref{eq-A9}) or $\Phi^{(v)}$ in eq.~(\ref{eq-A11}).  For this purpose, we see that the presence of $\delta({\bf r} - {\bf r}')$ and $\delta(t - t')$ on the RHS of eq.~(\ref{eq-A20}) suggests that the following results are true (they can be derived):
\begin{equation}
\grad' G({\bf r}, t |c|v| {\bf r}', t') = - \grad G({\bf r}, t |c|v| {\bf r}', t'),
\label{eq-A21}
\end{equation}
\begin{equation}
{\partial \over \partial t'} G({\bf r}, t |c|v| {\bf r}', t') = - {\partial \over \partial t}  G({\bf r}, t |c|v| {\bf r}', t).
\label{eq-A22}
\end{equation}

Then from eq.~(\ref{eq-A18}), we do the following (see Ref.~\cite{Yang-Nevels-v-gauge-2025}):
\begin{eqnarray}
{\partial \over \partial t} \left[ {c \over 4\pi} \left( {1 \over v^2} - {1 \over c^2} \right) {\partial \over \partial t} G({\bf r}, t |c|v| {\bf r}', t') \right]
\qquad \qquad \qquad \qquad \qquad \qquad \qquad \qquad \ \ \ 
\nonumber
\\ 
= {c \over 4\pi} \left\{ \left[ \left( \grad^2 - {1 \over c^2} {\partial^2 \over \partial t^2} \right)G({\bf r}, t |c|v| {\bf r}', t')\right] 
- \left[ \left( \grad'^2 - {1 \over v^2} {\partial^2 \over \partial t'^2} \right) G({\bf r}, t |c|v| {\bf r}', t') \right]  \right\}
\nonumber
\\
= {c \over 4\pi} \int \left[ \left( \grad^2 - {1 \over c^2} {\partial^2 \over \partial t^2} \right) 
  G({\bf r}, t |c| {\bf r}'', t'') \right] G({\bf r}'', t'' |v| {\bf r}', t') d^3r''dt''
\qquad \qquad \qquad \quad 
\nonumber
\\
  - {c \over 4\pi} \int 
  G({\bf r}, t |c| {\bf r}'', t'')  \left[ \left( \grad'^2 - {1 \over v^2} {\partial^2 \over \partial t'^2} \right) G({\bf r}'', t'' |v| {\bf r}', t') \right] d^3r''dt''
\qquad \qquad \quad 
\nonumber
\\
= {c \over 4\pi} \int \left[ - 4\pi \delta({\bf r} - {\bf r}'') \delta(t - t'') \right] G({\bf r}'', t'' |v| {\bf r}', t') d^3r''dt''
\qquad \qquad \quad \qquad \qquad \quad \ \ \,
\nonumber
\\  - {c \over 4\pi} \int 
  G({\bf r}, t |c| {\bf r}'', t'')  \left[ - 4\pi \delta({\bf r}'' - {\bf r}') \delta(t'' - t') \right] d^3r''dt''
\qquad \qquad \qquad \qquad \ \ \ \ 
\nonumber
\\
= - c G({\bf r}, t |v| {\bf r}', t') + c G({\bf r}, t |c| {\bf r}', t').
\qquad \qquad \qquad \qquad \qquad \qquad \qquad \qquad \qquad \quad \ 
\label{eq-A23}
\end{eqnarray}
Thus, ${\bf A}_{2}^{(v)}$ in eq.~(\ref{eq-A18}) becomes
\begin{eqnarray}
{\bf A}_2^{(v)}({\bf r},t) = c \grad \int dt \int \left[ G({\bf r}, t |c| {\bf r}', t') - G({\bf r}, t |v| {\bf r}', t') \right] \rho({\bf r}', t') d^3r'dt'
\nonumber
\\
= c \grad \int \left[ \Phi_c({\bf r}, t) - \Phi^{(v)}({\bf r}, t) \right]dt.
\qquad \qquad \qquad \qquad \qquad \ \ \ \,
\label{eq-A24}
\end{eqnarray}

For the Coulomb gauge, we simply set $v \to \infty$ in the above result to get,
\begin{equation}
{\bf A}_2^{(C)}({\bf r},t) = c \grad \int \left[ \Phi_c({\bf r}, t) - \Phi^{(C)}({\bf r}, t) \right]dt.
\label{eq-A25}
\end{equation}
See the Appendix for a derivation of eq.~(\ref{eq-A25}) strictly in the Coulomb gauge.

Our reason for converting the result from eq.~(\ref{eq-A18}) to (\ref{eq-A24}) is based on physics.  Any part of a wave propagating in a {\em uniform} medium will only propagate in one direction with just one speed. Infinite free space without boundary surfaces certainly qualifies as a uniform medium.  The result in eq.~(\ref{eq-A24}) fits this physical picture: there are two waves both propagating from ${\bf r}'$ to ${\bf r}$ but one with speed $c$ and the other with speed $v$.  But the result in eq.~(\ref{eq-A18}) paints a very different picture.  The wave starts from ${\bf r}'$, travels at speed $v$ to ${\bf r}''$, then changes its direction and goes from ${\bf r}''$ to ${\bf r}$ and propagates at a different speed $c$.  Moreover, the intermediate position ${\bf r}''$ extends to the entire space.  Because of these considerations, we prefer to use eq.~(\ref{eq-A24}) to represent ${\bf A}^{(v)}_2$.

\begin{flushleft}
{\bf Method 4}:
\end{flushleft}

Here, we derive ${\bf A}^{(v)}_2$ in eq.~(\ref{eq-A10}) by the method of gauge transformations (see also Ref.~\cite{Jackson-2002}).  We start from the potentials in the Lorenz gauge produced by the charge and current densities:
\begin{equation}
\Phi^{(L)} ({\bf r}, t) = \Phi_c ({\bf r}, t),
\qquad \qquad
{\bf A}^{(L)} ({\bf r}, t) = {\bf A}_c ({\bf r}, t).
\label{eq-A26}
\end{equation}
We define a gauge function $\chi$ by
\begin{equation}
\chi = c \int \left( \Phi_c - \Phi^{(v)} \right) dt. 
\label{eq-A27}
\end{equation}
After a gauge transformation on the Lorenz gauge with the gauge function $\chi$, we obtain the new potentials:
\begin{equation}
\Phi^{({\rm new})} = \Phi_c - {1 \over c} {\partial \chi \over \partial t} = \Phi^{(v)},
\label{eq-A28}
\end{equation}
\begin{equation}
{\bf A}^{({\rm new})} ={\bf A}^{(L)} + \grad \chi ={\bf A}_c + c \grad \int \left( \Phi_c - \Phi^{(v)} \right) dt. 
\label{eq-A29}
\end{equation}

In conclusion, we note that if one questions the legitimacy of eq.~(\ref{eq-A10}), then one is questioning the legitimacy of {\em all} of Methods 1 to 4.  These four methods all lead to eq.~(\ref{eq-A10}).  To the best of our abilities, so far we have {\em not} detected any illegal mathematical operations in these four methods proving that eq.~(\ref{eq-A10}) is a mathematically correct solution of eq.~(\ref{eq-A8}).

The most important and interesting result of Ref.~\cite{Yang-Nevels-2025} is expressed as follows (also observed by Onoochin~\cite{Onoochin-2025}).  Any potentials $({\bf A}, \Phi)$ that are solutions of Maxwell's equations for potentials always satisfy the following relation~\cite{note-5}:
\begin{equation}
{\bf A}({\bf r}, t) + c \grad \int \Phi({\bf r}, t) dt = {\bf A}_c({\bf r}, t) + c \grad \int \Phi_c({\bf r}, t) dt,
\label{eq-A30}
\end{equation}
for all ${\bf r}$ and $t$.  If we take $(-c^{-1} \partial / \partial t)$ and $(\grad \times)$ on both sides of the equation, we get
\begin{equation}
- {1 \over c} {\partial {\bf A}({\bf r}, t) \over \partial t} - \grad \Phi({\bf r}, t) 
= - {1 \over c} {\partial {\bf A}_c ({\bf r}, t) \over \partial t} - \grad \Phi_c ({\bf r}, t) = {\bf E}({\bf r}, t),
\label{eq-A31}
\end{equation}
\begin{equation}
\grad \times {\bf A}({\bf r}, t) = \grad \times  {\bf A}_c ({\bf r}, t) = {\bf B}({\bf r}, t).
\label{eq-A32}
\end{equation}
Eqs.~(\ref{eq-A30})-(\ref{eq-A32}) tell us that these potentials will always generate {\em gauge-invariant} electric and magnetic fields. All our potentials presented here and in Ref.~\cite{Yang-Nevels-2025} satisfy eq.~(\ref{eq-A30}), which means that all our potentials generate gauge-invariant fields.  If there are some potentials that differ in substance (not superficially in appearance) from our corresponding potentials ({\it e.g.}~other Coulomb-gauge potentials {\it vs.}~our Coulomb-gauge potentials), then the other potentials may run the risk of violating the principle of gauge invariance of the fields.  Gauge invariance is such an important principle of physics that it should be checked often and observed at all times.

\appendix
\section{Appendix: Derivation of eq.~(\ref{eq-A25}) in the Coulomb gauge}
\setcounter{equation}{0}
\renewcommand{\theequation}{A.\arabic{equation}} 

In the Coulomb gauge, eq.~(\ref{eq-A8}) becomes (setting $v \to \infty$)
\begin{equation}
\left( \grad^2 - {1 \over c^2} {\partial^2 \over \partial t^2} \right) {\bf A}^{(C)}_{2}
= {1 \over c} \left( {\partial \over \partial t} \grad \Phi^{(C)} \right).
\label{eq-app1}
\end{equation}
We solve the vector potential ${\bf A}^{(C)}_2$ by straightforward integrations:
\begin{eqnarray}
{\bf A}^{(C)}_2 ({\bf r}, t) = - {1 \over 4\pi c}   \grad {{\partial \over \partial t}}
\int G({\bf r}, t |c| {\bf r}'', t'') \Phi^{(C)}({\bf r}'', t'') d^3r''dt''
\qquad \qquad \quad \,
\nonumber
\\
= - {1 \over 4\pi c} \grad {{\partial \over \partial t}}
\int G({\bf r}, t |c| {\bf r}'', t'') G({\bf r}'' | {\bf r}') \rho({\bf r}', t'') d^3r''dt'' d^3r' ,
\quad \ \,
\label{eq-app2}
\end{eqnarray}
\begin{equation}
G({\bf r}'' | {\bf r}') = {1 \over | {\bf r}'' - {\bf r}'| }, \qquad
\grad''^2 G({\bf r}'' | {\bf r}') = - 4 \pi \delta({\bf r}'' - {\bf r}'),
\qquad \qquad \qquad \ 
\label{eq-app3}
\end{equation}
and the $c$-retarded Green's function $G({\bf r}, t |c| {\bf r}'', t'')$ is defined similarly to $G({\bf r}, t |v| {\bf r}', t')$ in eqs.~(\ref{eq-A12})-(\ref{eq-A13}).  We now differentiate eq.(\ref{eq-app2}) with respect to $t$ and use eqs.~(\ref{eq-A13}) and (\ref{eq-app3}) to get,
\begin{eqnarray}
{\partial {\bf A}^{(C)}_2 ({\bf r}, t) \over \partial t}
= - {c \over 4\pi} \grad \int \left[ \left( {1 \over c^2} {\partial^2 \over \partial t^2} \right) G({\bf r}, t |c| {\bf r}'', t'') \right] 
    G({\bf r}'' | {\bf r}') \rho({\bf r}', t'') d^3r''dt'' d^3r' 
 \qquad \ 
\nonumber
\\
= - {c \over 4\pi} \grad \int \left[4\pi \delta({\bf r} - {\bf r}'') \delta(t - t'') + \grad^2 G({\bf r}, t |c| {\bf r}'', t'') \right]
  G({\bf r}'' | {\bf r}') \rho({\bf r}', t'') d^3r''dt'' d^3r'
 \qquad 
  \nonumber
\\
= - c \grad \int G({\bf r} | {\bf r}') \rho({\bf r}', t) d^3r'
  - {c \over 4\pi} \grad \int \left[ \grad''^2 G({\bf r}, t |c| {\bf r}'', t'') \right]
  G({\bf r}'' | {\bf r}') \rho({\bf r}', t'') d^3r''dt'' d^3r'
  \nonumber
\\
= - c \grad \Phi^{(C)}({\bf r}, t) 
  - {c \over 4\pi} \grad \int  G({\bf r}, t |c| {\bf r}'', t'')
  \left[ \grad''^2 G({\bf r}'' | {\bf r}')  \right] \rho({\bf r}', t'') d^3r''dt'' d^3r'
  \qquad \qquad \ \ 
\nonumber
\\
= - c \grad \Phi^{(C)}({\bf r}, t) 
  - {c \over 4\pi} \grad \int  G({\bf r}, t |c| {\bf r}'', t'')
  \left[ - 4 \pi \delta({\bf r}'' - {\bf r}')  \right] \rho({\bf r}', t'') d^3r''dt'' d^3r'
\qquad \qquad 
\nonumber
\\
= - c \grad \Phi^{(C)}({\bf r}, t) 
  + {c} \grad \int  G({\bf r}, t |c| {\bf r}'', t'') \rho({\bf r}'', t'') d^3r''dt''
\qquad \qquad \qquad \qquad \qquad \qquad \quad \ \, 
\nonumber
\\
= - c \grad \Phi^{(C)}({\bf r}, t) + c \grad \Phi_c ({\bf r}, t).
\qquad \qquad \qquad \qquad \qquad \qquad \qquad \qquad \qquad \qquad \qquad \qquad \, 
\label{eq-app4}
\end{eqnarray}
Thus, we have derived eq.~(\ref{eq-A25}) strictly in the Coulomb gauge.

\end{document}